\documentclass[10pt,conference]{IEEEtran}
\usepackage{lipsum,multicol}
\usepackage{color}
\newcommand{\Ex}{\mathop{\bf E\/}}

\usepackage{cite}

\usepackage[cmex10]{amsmath}
\usepackage{cases}  % for begin numcases

\usepackage{algorithmic}

\usepackage{array}

\usepackage{mdwmath}
\usepackage{mdwtab}

\usepackage{epstopdf}

\ifCLASSINFOpdf
   \usepackage[pdftex]{graphicx}
\else
\fi

\usepackage{subfigure}

\usepackage{url}
\begin{document}
%

% **************** PAPER TITLE  *************************

%\title{On The Flow-level Delay of MIMO Wireless Channels}
\title{On the Flow-Level Delay of a Spatial Multiplexing MIMO Wireless Channel}

% can use linebreaks \\ within to get better formatting as desired

% **************** EOF PAPER TITLE  *************************

% ********************  AUTHOR NAMES AND AFFILIATIONS ************************

\author{\IEEEauthorblockN{Kashif Mahmood~\IEEEauthorrefmark{1}, Amr Rizk~\IEEEauthorrefmark{7}, Yuming Jiang~\IEEEauthorrefmark{1}}       % <-this % stops a space

\IEEEauthorblockA{\IEEEauthorrefmark{1} Centre for Quantifiable Quality of Service in Communication Systems,~NTNU,~Norway\\
\IEEEauthorrefmark{7} Institute of Communications Technology, Leibniz Universit\"{a}t Hannover, Germany }}
%\author{\IEEEauthorblockN{abc}
%\author{\IEEEauthorblockN{a,b,c}
        % <-this % stops a space
%\IEEEauthorblockA{Centre for Quantifiable Quality of Service in Communication Systems,~NTNU,~Norway\\

% ******************** use for special paper notices  *************************

%\IEEEspecialpapernotice{(Invited Paper)}

% ****************** MAKE THE TITLE AREA  ***************************************
\maketitle

% ****************** EOF MAKE THE TITLE AREA  ***************************************

% ****************** ABSTRACT  ***************************************

\begin{abstract}
%\boldmath
The MIMO wireless channel offers a rich ground for quality of service analysis. In this work, we present a stochastic network calculus analysis of a MIMO system, operating in spatial multiplexing mode, using moment generating functions (MGF). We quantify the spatial multiplexing gain, achieved through multiple antennas, for flow level quality of service (QoS) performance. Specifically we use Gilbert-Elliot model to describe individual spatial paths between the antenna pairs and model the whole channel by an $N$-State Markov Chain, where $N$ depends upon the degrees of freedom available in the MIMO system. We derive probabilistic delay bounds for the system and show the impact of increasing the number of antennas on the delay bounds under various conditions, such as channel burstiness, signal strength and fading speed. Further we present results for multi-hop scenarios under statistical independence.

\end{abstract}

% IEEEtran.cls defaults to using nonbold math in the Abstract which is common
% If you want to do the uncommon use command \boldmath at the very start of the abstract to achieve this.

% ****************** EOF ABSTRACT  ***************************************
% no keywords

%\begin{IEEEkeywords}
%Stochastic Network Calculus, multiple-input-multiple-output (MIMO), Markov modeling, moment generating function (MGF).
%\end{IEEEkeywords}

% ****************** PEER REVIEW  ***************************************

% For peer review papers, you can put extra information on the cover page as needed: \ifCLASSOPTIONpeerreview
% \begin{center} \bfseries EDICS Category: 3-BBND \end{center}
% \fi

% For peerreview papers, this IEEEtran command inserts a page break and creates the second title. It will be ignored for other modes.

\IEEEpeerreviewmaketitle

% ****************** EOF PEER REVIEW  ***************************************

\section{Introduction}
% no \IEEEPARstart
The last decade has seen an explosive growth in wireless network services and consequently an increasing demand of high data rates on wireless networks. 
Multiple-input-multiple-output (MIMO)~\cite{MIMO:Foschini:98:limits,MIMO:Teletar99:CapacityOfMultiAntennaGaus} communication has generated a lot of interest in recent years for providing such high data rates. It offers enormous capacity gains as compared to traditional single input single output (SISO) communication~\cite{WirelessMIMO:Paulraj2003:introductionToST}. The importance of MIMO technology can be seen from the fact that IEEE $802.11$ is actively pursuing $802.11$n which promises high throughput using multiple antennas. To date, a significant amount of work on MIMO has been done in quantifying its capacity limits under different channel conditions, antenna settings, etc., using classical information theory under the assumption of infinite delay~\cite{Wireless:Tse05:FundamentalsOfWireless:Bk}. This assumption has been working well for delay insensitive applications and for channels without memory.

However, recently there has been a rapid growth of delay sensitive applications such as VoIP and IP video, which are also run on wireless networks where MIMO is being deployed. In addition, MIMO links have been considered for delay critical applications such as MANETS~\cite{MIMO:Adhoc:Chen06}, for they overcome the problems of low system throughput. Hence, it is also essential to provide quality of service (QoS) (such as delay\footnote{Throughout the paper the term \emph{delay} refers to \emph{queuing delay}.} bounds) on MIMO links in addition to high data rates. For this, a critical issue is to analyze their flow level performance. 

The focus of this paper is on flow-level delay analysis of a spatial multiplexing MIMO wireless channel. Specifically, we derive probabilistic delay bounds for the channel and show the impact of increasing the number of antennas on the delay bounds under various conditions, such as channel burstiness, signal strength and fading speed. The theory, which the analysis is based on, is moment generating function (MGF) based stochastic network calculus. In stochastic network calculus, a communication system is described by the arrivals at the ingress, the service it provides and the departures at the egress. The stochastic network calculus framework provides the freedom of a general separation in the characterization of arrivals and service, which allows the combination of different arrival and service models. It enables the derivation of flow level performance bounds such as $\text{Prob}[Delay > d ] \le \varepsilon$, i.e. the probability of the delay exceeding a certain threshold $d$ is upper bounded by $\varepsilon$. It also incorporates the ability to characterize multiplexing and demultiplexing of flows and the derivation of end-to-end performance bounds through composition \cite{ciucu06,NetCal:Fidler2006:EoEProbabNetCalWithMGF}. The MGF based stochastic network calculus, as proposed in \cite{NetCal:Chang00:PerGuaran:Bk}, established and developed in \cite{NetCal:Fidler2006:EoEProbabNetCalWithMGF}, enables more easily the consideration of statistical independence between arrivals and system service and between the service provided at consecutive hops. A comprehensive view of stochastic network calculus can be found in \cite{NetCal:Jiang08:SNC:Bk}, \cite{fidler_survy_2010}. 

In the literature, many works, which are related to QoS analysis of wireless channels, can be found. They include link-layer channel modeling \cite{WirelessMarkov:Sadeghi08:FiniteStateMarkov:Survey}, where the focus has been on developing models for channels with memory, which capture the bursty nature of errors on wireless channels. Effective bandwidth theory has been employed to analyze wireless channels \cite{EBWireless:AMohammadi97}. The concept of effective capacity, which is directly related to effective bandwidth, was developed in \cite{NetCal:Wu06:EffecCap} for wireless channel QoS analysis. By using large deviations techniques a similar approach was developed in~\cite{EBWireless:Li07} to relate effective bandwidth theory to channel capacity and derive delay bounds in conjunction with Markov modeling. In addition, {\em stochastic network calculus}, which as a more general theory covers effective bandwidth and effective capacity, has recently been applied for QoS analysis of wireless channels \cite{NetCal:Jiang05:StochServGuarantServerModel,NetCal:Fidler06:MGFfadingChannel,NetCal:Verticale09:Wireless:ClosedFormExpressionForDelayMVABound}. However, none of these works analyzed MIMO channels. 

There are two transmission techniques for MIMO systems. \emph{Spatial multiplexing} is used to increase throughput by transmitting \emph{independent} data flows over different antennas. \emph{Spatial diversity}, on the other hand, increases reliability by transmitting \emph{redundant} flows of information in parallel. There has been some works reported for MIMO diversity systems~\cite{EBWireless:Tang2007:CrossLayerModelingforQoS,MIMO:Diversity:Zorzi99:Lateness}. In \cite{EBWireless:Tang2007:CrossLayerModelingforQoS}, the concept of effective capacity has been used to carry out QoS analysis of MIMO diversity systems using finite state Markov chains to describe the \emph{received signal to noise ratio} (SNR). However, to the best of our knowledge, QoS analysis of MIMO systems in spatial multiplexing mode is still missing. 
In addition, it is known that, besides average received SNR, the MIMO \emph{channel matrix}, which describes the scattering environment, has a profound effect on channel capacity~\cite{WirelessMIMO:Paulraj2003:introductionToST} and hence the delay.

The contribution of this work is as follows. (1) We present QoS analysis of a MIMO system under \emph{spatial multiplexing} mode. In the analysis, we incorporate statistical independence of arriving traffic and the service provided by the channel. We also present results for end-to-end performance in a multi-hop scenario, where statistical independence of consecutive hops holds. (2) The \emph{effect of individual channel matrix elements} on delay bounds is studied keeping the received SNR constant. (3) We provide a method of aggregating the states based on degrees of freedom available in the MIMO system, which results in \emph{reduction of the state space} from $2^{L^2}$ to $L+1$.
%
%\begin{itemize}
%\item We present QoS analysis of a MIMO system under \emph{spatial multiplexing} mode. In the analysis, we incorporate statistical independence of arriving traffic and the service provided by the channel. We also present results for end-to-end performance in a multi-hop scenario, where statistical independence of consecutive hops holds.
%\item The \emph{effect of individual channel matrix elements} on delay bounds is studied keeping the received SNR constant.
%\item We provide a method of aggregating the states based on degrees of freedom available in the MIMO system, which results in \emph{reduction of the state space} from $2^{L^2}$ to $L+1$.
%\end{itemize}
%This makes our methodology applicable to a wide range of physical environments from rich scattering environment to a sparse scattering environment.

The paper is organized as follows. We first present the system model and the assumptions in Sec.~\ref{sec:prelim}, and the necessary background on stochastic network calculus and MIMO in Sec.~\ref{Subsec:NetCal} and Sec.~\ref{Subsec:MIMO} respectively. We then carry out the analysis in Sec.~\ref{Sec:ProbForm}. Numerical results for a parameterization according to IEEE $802.11$n are presented in Sec.~\ref{Sec:Results}.% to exemplify the proposed methodology.
%The paper is organized as follows. We first present the system model of Markov Modeling for MIMO systems and the assumptions in Section~\ref{sec:prelim}. The necessary background on stochastic network calculus and MIMO in Sections~\ref{Subsec:NetCal} and ~\ref{Subsec:MIMO} respectively. We then carry out the system analysis in Section~\ref{Sec:ProbForm}. Numerical results for a parameterization according to IEEE $802.11$n are presented in Section~\ref{Sec:Results} to exemplify the proposed methodology.
%
\section{System Model, Assumptions and Preliminaries}
\label{sec:prelim}
We consider the single user MIMO case of a transmitter and a receiver with $N$ and $M$ antennas respectively. The MIMO system adopts spatial multiplexing. 
We assume a discrete time block fading model in which the channel is constant over the time length of one data block.
We introduce a service model based on Gilbert-Elliott (GE) channels~\cite{WirelessMarkov:Gilbert60:CapacityBurstNoiseChannel, elliot_noise}.
Each fading channel is modeled by two states, \{good\} and \{bad\}, where a good channel is assumed for an error free transmission while a bad channel represents an unsuccessful transmission.
We use MGF stochastic network calculus (NetCal) to carry out QoS analysis of the above mentioned MIMO system. 
%Before we present the analysis, we first present the preliminaries of MGF Stochastic NetCal and MIMO system.
\subsection{MGF Stochastic Network Calculus}\label{Subsec:NetCal}
Stochastic network calculus is a framework for QoS analysis and flow level performance evaluation of communication systems. This framework has been firstly used for characterizing fading channels in~\cite{NetCal:Jiang05:StochServGuarantServerModel}. We resort to a stochastic NetCal approach based on moment generating functions, introduced in~\cite{NetCal:Fidler2006:EoEProbabNetCalWithMGF}, for the analysis of the considered MIMO wireless fading channel. The first analysis of block fading channels using the MGF based stochastic NetCal was carried out in \cite{NetCal:Fidler06:MGFfadingChannel}. We carry forward the properties of the stochastic network calculus approach to MIMO wireless channels.

The stochastic NetCal with MGFs allows efficiently exploiting statistical independence of arrivals and service processes and the statistical independence of service provided at concatenated hops. It enables the analysis of a wide variety of scenarios for different arrival and service processes where the MGF can be calculated.

We build on the dynamic server definition from \cite{NetCal:Chang00:PerGuaran:Bk} to model the time varying capacity of the channel. The discrete time arrivals $A(0,t)$ and service $S(s,t)$ are assumed to be statistically independent and stationary random processes. $S(s,t)$ is the service and $A(s,t)$ denote the cumulative arrival data provided in the interval $(s,t]$ respectively. The MGF of a stationary process $X(t)$ is given by
\begin{equation}
\mathsf{M}_X(\theta,t) = \Ex \left[e^{\theta X(t)}\right]
\end{equation}
while $\overline{\mathsf{M}}_X(\theta,t) = \mathsf{M}_X(-\theta,t)$ for parameter $\theta > 0$, $t \geq 0$ and $\Ex \left[\cdot\right]$ denotes the expectation. For a set of useful MGF properties we refer the interested reader to~\cite{EB:Kelly96,NetCal:Fidler2006:EoEProbabNetCalWithMGF}. Using the MGF stochastic network calculus performance bounds on backlog and delay were generally derived in~\cite{NetCal:Fidler2006:EoEProbabNetCalWithMGF} from Chernoff's theorem for arrivals and service random process that posses the MGFs $\mathsf{M}_A(\theta,t)$ and $\overline{\mathsf{M}}_S(\theta,t)$ respectively. One of the strengths of stochastic NetCal lies in the characterization of end-to-end service by means of composition of the service provided at each hop, i.e. finding a single dynamic server characterization equivalent to multiple dynamic servers in series. For $\eta$ servers in tandem providing stochastically independent service, the MGF of the end-to-end system is bounded by [Th. 1 from \cite{NetCal:Fidler2006:EoEProbabNetCalWithMGF}] for $t\geq0$
\begin{equation}
\overline {\mathsf{M}}_{S_{e2e}}(\theta,t) \leq \left(\overline {\mathsf{M}}_{S_{1}}(\theta) \ast \dots \ast\overline {\mathsf{M}}_{S_{\eta}}(\theta)\right)(t)
\label{eq:multihop}
\end{equation}
where $\ast$ is the convolution operator known from system theory.

A stochastic bound on the delay $W$ such that $P\left[W>d\right] < \varepsilon$ assuming FIFO scheduling is derived in \cite{NetCal:Fidler2006:EoEProbabNetCalWithMGF}, with
%
%\begin{equation*}
%\label{E:backlogBound}
%b=\inf_{\theta > 0} \left [  \frac{1}{\theta}  \left (\ln \sum_{s=0}^{\infty} \mathsf{M}_A(\theta,s) \overline {\mathsf{M}}_S(\theta,s)  - \ln \varepsilon        \right ) \right ]
%\end{equation*}
%
\begin{equation*}
\label{E:DelayBound}
d\!=\! \inf_{\theta > 0} \! \left [ \! \inf \! \left [ \! \tau :  \frac{1}{\theta}  \left ( \! \ln \! \sum_{s=\tau}^{\infty} \! \mathsf{M}_A(\theta,s- \tau) \overline{\mathsf{M}}_S(\theta,s) \! - \! \ln \varepsilon \right ) \leq 0 \! \right ] \! \right ].
\end{equation*}

%In this paper we target finding stochastic delay bounds for both single and multi-hop MIMO fading channels. We model the wireless channel such that a calculation of the MGF $ \overline {\mathsf{M}}_S(\theta,t) $ of the service is possible. We show results for delay bounds as described by the formulae above for a variety of scenarios.
%
\subsection{MIMO with Spatial Multiplexing}\label{Subsec:MIMO}
%The physical layer model in this work comprises of MIMO system.
%Consider a narrow-band single user MIMO system with $N$ transmit and $M$ receive antennas.
The $M \times 1$ received signal vector $\mathbf{y}$ of a MIMO system with $N$ transmit and $M$ receive antennas is  given as
\[
\mathbf{y} = \sqrt{\frac{\rho}{N}} \mathbf{Hs}+\mathbf{n}
\]
where $\mathbf{s}$ is a $N \times 1$ transmitted signal vector,
while $\mathbf{n}$ is a $M \times 1$ zero mean, complex Gaussian noise with independent and identically distributed real and imaginary parts and $\rho$ is the average SNR at the receiver.
We further assume $\Ex[\mathbf{nn}^\dag]=N_o\mathbf{I_M}$ i.e. the noise corrupting the different receivers are independent where $[\cdot]^\dag$ denotes the matrix conjugate transpose, $N_o$ is the noise power density and $\mathbf{I_M}$ is the identity matrix of dimension (M).
$\mathbf{H}$ is $M \times N$ MIMO channel matrix given as
\[
\mathbf{H}=
\begin{bmatrix}
h_{1,1} & h_{1,2} & \cdots & h_{1,N}\\
h_{2,1} & h_{2,2} & \cdots & h_{2,N}\\
\vdots & \vdots  & \ddots & \vdots\\
h_{M,1} & h_{M,2} & \cdots & h_{M,N}\\
\end{bmatrix}
 \]
where $h_{m,n}$ represents the fading path gain from the $n$th transmit antenna to the $m$th receive antenna.
We use the finite scatterers MIMO channel model~\cite{MIMO:Burr03:FiniteScatterModel}, which is widely used for describing multipath environments, to describe the individual channel gains. Under this model, the channel path gain~$h_{m,n}$, assuming no line of sight (NLOS) component, is given as~\cite{MIMO:Uthansakul10:InvestigationsIntoMIMOCap:FiniteScattererModel}
\begin{equation}
\label{E:PathGainFiniteScatterer}
h_{m,n}=\sum_{s=1}^{N_s} \alpha_s A_s \exp \left [-j \frac{2\pi}{\lambda} ( D_{n,s}+D_ {s,m})   \right]
\end{equation}
where $N_s$ is the number of scatterers, $\alpha_s$ is a uniformly distributed complex scattering coefficient for a scatterer $s$.
$D_{n,s}$ is the distance between the transmit antenna $n$ and the scatterer $s$ while $D_{s,m}$ is the distance between the scatterer $s$ and the receive antenna $m$ and $A_s$ is the attenuation constant of the communication path.

The squared Frobenius norm of $\mathbf{H},\Vert\mathbf{H}\Vert_F^2$ represents the total power gain of the channel and is given as
\[
\Vert\mathbf{H}\Vert_F^2 = \sum_{m=1}^{M} \sum_{n=1}^{N} \vert h_{m,n}\vert^2
\]
The meaning of the SNR $\rho$ depends upon the normalization of the channel matrix\cite{MIMO:Loyka:Normalization}.
We apply the most commonly used normalization~\cite{MIMO:Teletar99:CapacityOfMultiAntennaGaus}
\[
\Vert\mathbf{H}\Vert_F^2 = N\cdot M
\]
In this case, $\rho$ is the SNR per receiver from all transmitters while $\rho/N$ is the SNR per receiver per transmitter.

The MIMO capacity under a given realization of channel matrix for uncorrelated equal power sources is given as~\cite{MIMO:Teletar99:CapacityOfMultiAntennaGaus}
\begin{equation}\label{E:MIMOEPCapFormulalogDet}
C =  log_2 \left [det \left ( I_M + \frac {\rho}{N} \mathbf{HH^{\dag}}   \right )     \right ].
\end{equation}

MIMO offers spatial multiplexing gain in which independent data flows are transmitted over different antennas to maximize the data rate.
Spatial multiplexing takes advantage of the extra Degrees Of Freedom (DOF) made available due to the multiple antennas.
A MIMO channel has up to $min(N,M)$ DOF \cite{MIMO:Foschini:98:limits}, \cite{MIMO:Teletar99:CapacityOfMultiAntennaGaus}.
In this work, we quantify the spatial multiplexing gain by modeling the available DOF in a MIMO system using a discrete time Markov Chain.
The available DOF, among others, depends upon the physical environment in which the MIMO system operates, which is represented by the channel matrix elements.
We next formulate the problem for QoS analysis of MIMO system for the general case in which entries of $H$ are formulated as in (\ref{E:PathGainFiniteScatterer}).

%\section{MIMO analyzed by Degree of Freedom}\label{Sec:ProbForm}
\section{The Analysis}\label{Sec:ProbForm}

The discrete time block fading model of a MIMO system with $N$ and $M$ transmit and receive antennas respectively can be represented by a discrete time  Markov chain. An immediate idea is to use $2^{N \times M}$ states to describe the system. However, it introduces exponential explosion of the state space and consequently the computation complexity. %In addition, due to inherent dependence among such states due to properties of spatial multiplexing in MIMO, it is difficult to link each state to a transmission rate as to be discussed below. 

In our analysis, we define states for the Markov chain based on the concept of degree of freedom (DOF). Particularly, for the $N \times M$ MIMO, $K=min(N,M) + 1$ states are adopted, where the $i$th state denotes that there are $(i-1)$ DOF available in the channel and has a stationary state distribution vector $\boldsymbol{\pi}$.

\begin{figure}[tb]
\centering
\includegraphics[width=0.4\textwidth]{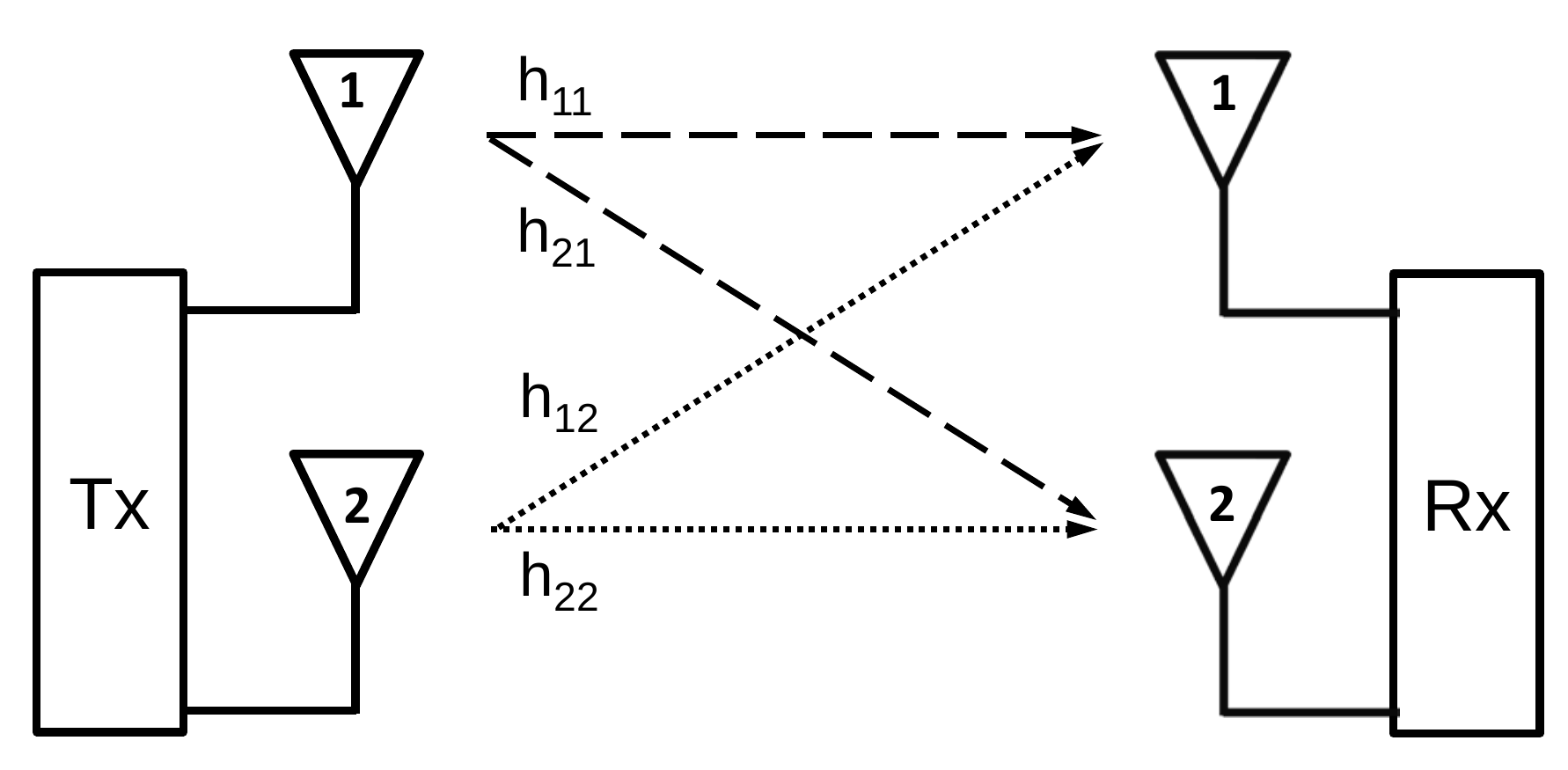}
\caption{MIMO with Spatial Multiplexing Mode}
\label{F:MIMO2by2}
\end{figure}

%The discrete time block fading model of a MIMO system with $N$ and $M$ transmit and receive antennas respectively can be represented by a discrete time multi-dimensional Markov chain with $K=min(N,M) + 1$ states, where the $i$th state denotes the degree of freedom $(i-1)$ available in the channel and has a stationary state distribution vector $\boldsymbol{\pi}$.
The Markov chain evolves according to the states of the underlying spatial paths, which dictate the available DOF.
These spatial paths form the sub-states of the Markov Chain. %The GE model is used to describe the underlying spatial paths.
A particular path~$h_{m,n}$ can be either \{good\} or \{bad\} (abbreviated g resp. b) under the GE model, depending upon its power which is in turn dictated by the number of scatterers~$N_s$ in~\eqref{E:PathGainFiniteScatterer}.
The $2\times 2$ scenario including four spatial paths $\left\{h_{11},h_{12},h_{21},h_{22}\right\}$ is depicted in Fig.~\ref{F:MIMO2by2}.
A combination \{gggg\} denotes that all the paths are in the {good} state.

We assume that the path stays in the good state with probability $1-p_{gb}$ and in the bad state with probability $1-p_{bg}$.
The probability to go from the good state to the bad state is therefore $p_{gb}$ and from the bad state to the good state is $p_{bg}$.
For a certain path the probability of being in the \{bad\} state, i.e. the block error probability for that path, is given as
\begin{equation}
\omega=\frac{p_{gb} } { p_{gb}+p_{bg} }.
\label{eq:block_error_prob}
\end{equation}

It is worth highlighting that for spatial multiplexing MIMO, whenever two paths targeting the same receiver antenna are \{good\}, such as \{ggbb\}, the receive antenna can only decode one of them. In addition, if more than one path with the same transmitter are \{good\}, such as \{gbgb\}, the transmission can be counted only once. Note that we are not analyzing automatic repeat request (ARQ) behavior. We assume that whenever all paths from a transmit antenna are \{bad\} the data blocks intended to that antenna are redirected to the other antennas.% such that no reordering at the receiver is needed as the data blocks arrive in a FIFO manner.

%It is worth highlighting that for spatial multiplexing MIMO, whenever two paths targeting the same receiver antenna are \{good\}, such as \{ggbb\}, the receive antenna can only decode one of them. In addition, if more than one paths with the same transmitter are \{good\}, such as \{gbgb\}, the transmission can be counted only once. Due to these strong correlations, if the $2^{N \times M}$ state Markov chain had been adopted, it would be difficult to decide the transmission rate of each state. Note that we are not analyzing automatic repeat request (ARQ) behavior. We assume that whenever all paths from a transmit antenna are \{bad\} the data blocks intended to that antenna are redirected to the other antennas.% such that no reordering at the receiver is needed as the data blocks arrive in a FIFO manner.

For the Markov chain whose state is based on DOF, the transmission rate~$r_i$ in state~$i$ is given as
\begin{equation}
r_i=min\{\overline {C}_{s,i} \}
\label{eq:r_i}
\end{equation}
where $\overline {C}_{s,i}$ is the average capacity of sub-state~$s$ of state~$i$ obtained by taking the average of $C$ in~\eqref{E:MIMOEPCapFormulalogDet}.
Here, $r_i$ can be considered as the minimum capacity achievable in the state $i$, which holds in the context of delay limited capacity~\cite{MIMO:OutageCap:NonZero:biglieri2001:limiting}.
%The averaging for a given realization is performed over the different c

The rates $r_i$ are ordered in the diagonal matrix $\mathbf{R}(\theta)$ as $\text{diag}(e^{\theta r_1},\dots,e^{\theta r_K})$.
The first order capacity is given as $\sum_{i=1}^K r_i\pi_i$.
The transition probability matrix is represented as $\mathbf{Q}$, where its element $p_{ij}$ denotes the transition probability of state $i$ to state $j$.
The MGF of the random process $S$ described by such Markov chain is given for $\theta > 0$ and $t \geq 0$ by~\cite{NetCal:Chang00:PerGuaran:Bk}
\begin{equation}
\label{E:MS}
\overline{\mathsf{M}}_S(\theta,t) = \boldsymbol{\pi} (\mathbf{R}(-\theta)\mathbf{Q})^{t-1}\mathbf{R}(-\theta)\mathbf{1}
\end{equation}
where $\mathbf{1}$ is column vector of ones.

Next we consider the simplest case of $2$x$2$ MIMO operating in spatial multiplexing mode.
A $2$x$2$ MIMO has $2$ degrees of freedom and is hence described by a three state Markov  chain:
\begin {itemize}
\item State $1$: DOF$=0$.
\item State $2$: DOF$=1$.
\item State $3$: DOF$=2$.
\end {itemize}

\begin{figure}[t]
\centering
\includegraphics[width=0.9\columnwidth]{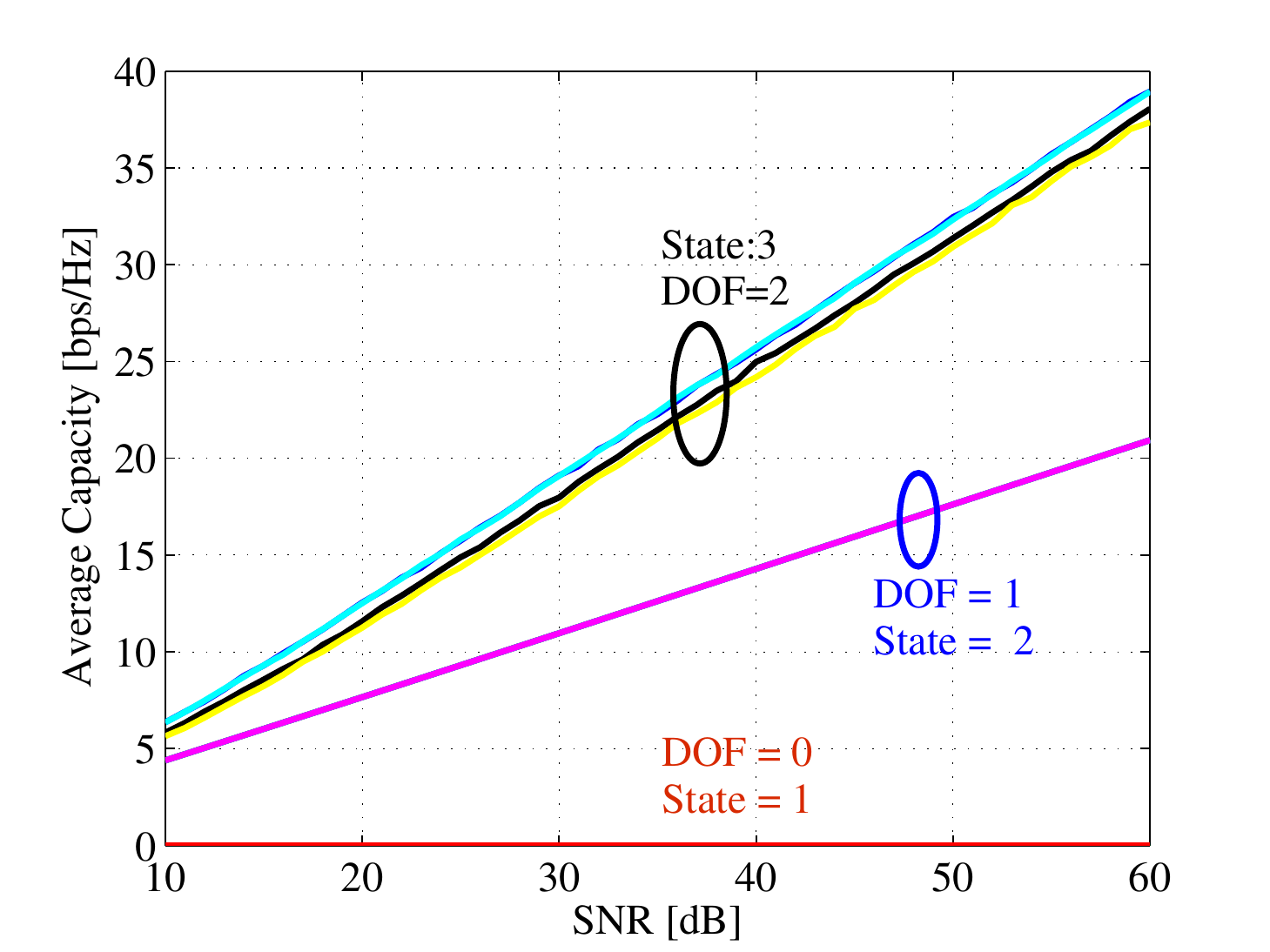}
\caption{States of the Markov Chain for N = 2}
\label{F:cap_substates}
\end{figure}

Each state is an aggregation of the sub-states where the aggregation is performed on the basis of DOF. Fig.~\ref{F:cap_substates} depicts this aggregation. The rate $r_i$ supported by the state $i$ with DOF $i-1$ is the minimum rate achievable in this state. 
\begin {itemize}
\item State $1$: bbbb.
\item State $2$: ggbb, bbgg, gbgb, bgbg, gbbb, bgbb, bbgb, bbbg.
\item State $3$: gggg, gggb, ggbg, gbgg, bggg, gbbg, bggb.
\end {itemize}
The transition matrix $\mathbf{Q}$ of this three state process can be calculated using conditional probabilities and the law of total probability.
The steady state vector follows accordingly using $\boldsymbol{\pi}~=~\boldsymbol{\pi} \cdot \mathbf{Q}$, which is given here as
\begin{eqnarray*}
\boldsymbol{\pi}^{1\times3} = [\omega^4~~4\omega^2(1-\omega)^2 + 4\omega^3(1-\omega)~~\\ (1-\omega)^4 + 4\omega(1-\omega)^3 + 2\omega^2(1-\omega)^2 ]
\end{eqnarray*}
The corresponding rate matrix $\mathbf{R}$ is given as $\mathbf{R}(\theta)=\text{diag}(1,e^{\theta r_2},e^{\theta r_3})$, where we assume that there is no workload processed in state~$1$. The method is easily extended to $N$x$N$ MIMO systems. 

With (\ref{E:MS}) decided, delay bounds for the MIMO system are readily obtained using stochastic network calculus results discussed in Sec \ref{Subsec:NetCal}. 

%In the next section, we present numerical results for $N$x$N$ with $N\in\{2,3,4\}$.

\section{Numerical Results}
\label{Sec:Results}

In this section we present numerical results for the $N$x$N$ MIMO channels as described in the previous section for different $N$. The delay bounds presented are calculated numerically based on the formulae in Sec \ref{Subsec:NetCal}. We use a parameterization for the channel and for the traffic source to compute the numerical results for a NLOS scenario.

We parameterize the MIMO channel according to 802.11n using $40$ MHz channels~\cite{MIMO:80211n:standard}. We choose a base time unit for the discrete time model of $31\mu s$ which is approximately the time needed to transmit a data block of $2312$ Bytes over a $600$ Mbps channel, the maximum rate given in~\cite{MIMO:80211n:standard}. As we consider scenarios with $N$x$N$ antennas we obtain the capacity of each of the DOF states through an extensive simulation as described in section~\ref{Subsec:MIMO}. We omit showing confidence intervals as these turn out to be very small.
We set $N_s=500$ to describe a good channel which employs that the NLOS component power is enough to support error free transmission while a bad channel is described by zero NLOS component power. We set $A_s=1$ $\forall s$ while the free space attenuation factor $1/(D_{n,s}D_ {s,m})$ is taken to be a constant such as in \cite{MIMO:Uthansakul10:InvestigationsIntoMIMOCap:FiniteScattererModel}.
For the GE model, we set the transition probability from \{bad\} to \{good\} $p_{bg} = 0.1$ and the probability from \{good\} to \{bad\} $p_{gb} = 0.01$ similar to \cite{NetCal:Fidler06:MGFfadingChannel}.

For the discrete time model we use a periodic source that generates arrival traffic. Such traffic source with period $\tau$ produces $\sigma$ units of workload (data blocks) at times $\left\{U \tau + n \tau, n = 0,1,\ldots\right\}$ where $U$ is the starting time which is uniformly distributed  in the interval $[0,1]$. The MGF of such a source is known (see e.g.~\cite{EB:Kelly96}) as
\[
\mathsf{M}_{A}(\theta,t) = e^{\theta \sigma \left\lfloor \frac{t}{\tau}\right\rfloor} \left(1+\left(\frac{t}{\tau} - \left\lfloor \frac{t}{\tau}\right\rfloor\right)\left(e^{\theta \sigma} - 1\right)\right)
\]
for $t \geq 0$ and $\theta > 0$. The framework described before enables the derivation of probabilistic performance bounds, such as for the delay, for a various number of traffic sources whenever the MGF exists. A collection of such MGFs for sources of different statistical properties can be found in \cite{EB:Kelly96} and references therein. In this work we focus on the periodic source model and exploit the enhancement in QoS, in terms of delay bound, delivered by the use of multiple antennas at the transmitter and receiver. The periodic source can be seen as a streaming source, generating $\sigma$ data blocks every time period $\tau$, which enables setting the arrival rate $v$. The source arrival rates are tuned to meet modulation and coding schemes (MCS) from the IEEE 802.11n standard \cite{MIMO:80211n:standard}, i.e. rate 240 Mbps corresponds to MCS-13 for $N = 2$ antennas. We parameterize the source such that its period $\tau$ is 10 time units and set the number of generated data blocks $\sigma$ to achieve a given arrival rate $v$. We fix $\varepsilon = 10^{-6}$ and $v = 240$ Mbps unless stated otherwise.

\begin{figure}[t]
	\centering
		\includegraphics[width=0.90\columnwidth]{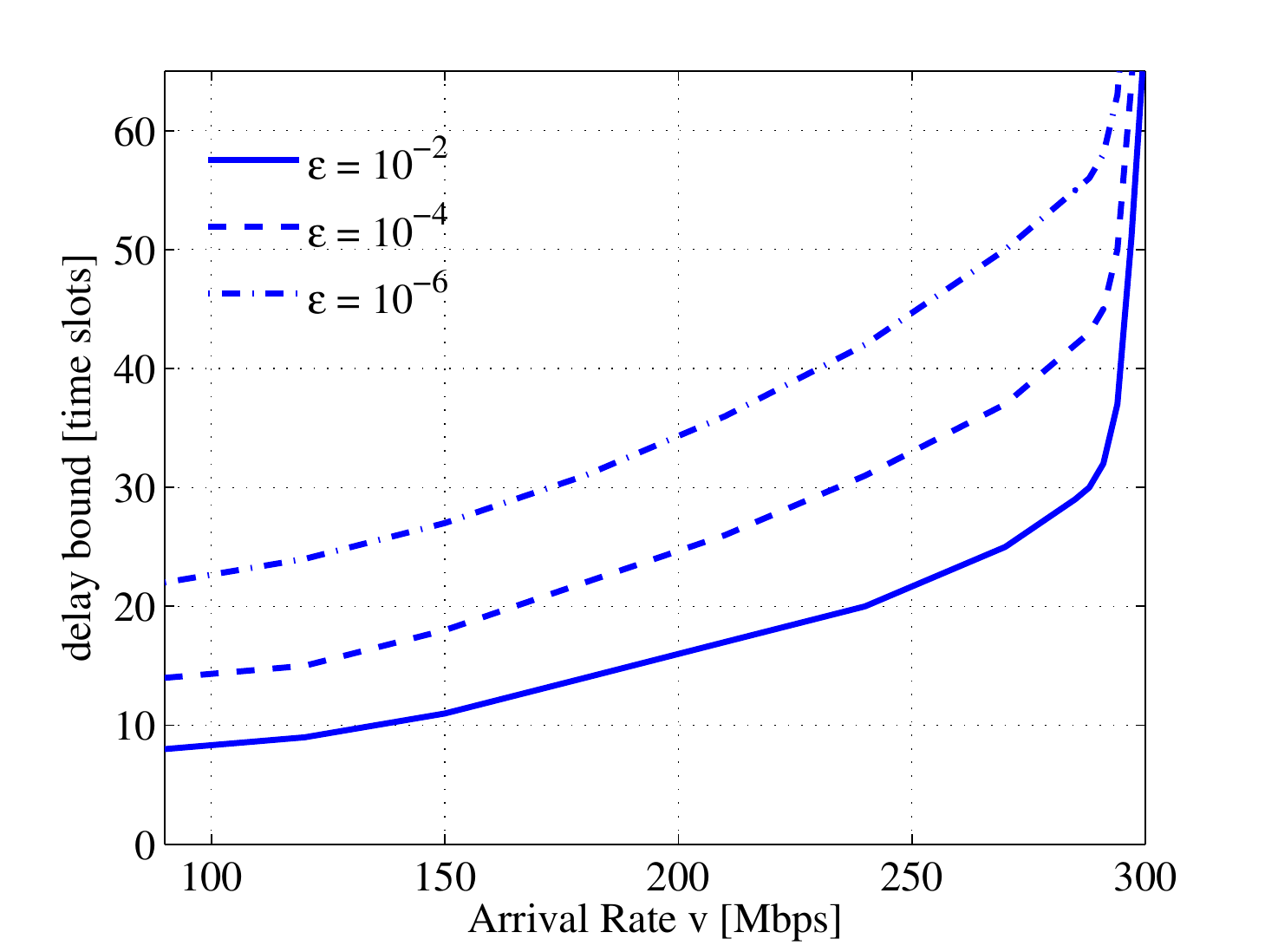}
	\caption{Stochastic delay bounds for N = 2}
	\label{fig:dly_bnd_arr_rate_N2}
\end{figure}

Fig.~\ref{fig:dly_bnd_arr_rate_N2} depicts numerical results on the stochastic delay bound as a function of the arrival traffic rate $v$ for different violation probabilities $\varepsilon$. Observe the degradation in terms of delay bound as the violation probability gets tighter.
\begin{figure}[t]
	\centering
		\includegraphics[width=0.90\columnwidth]{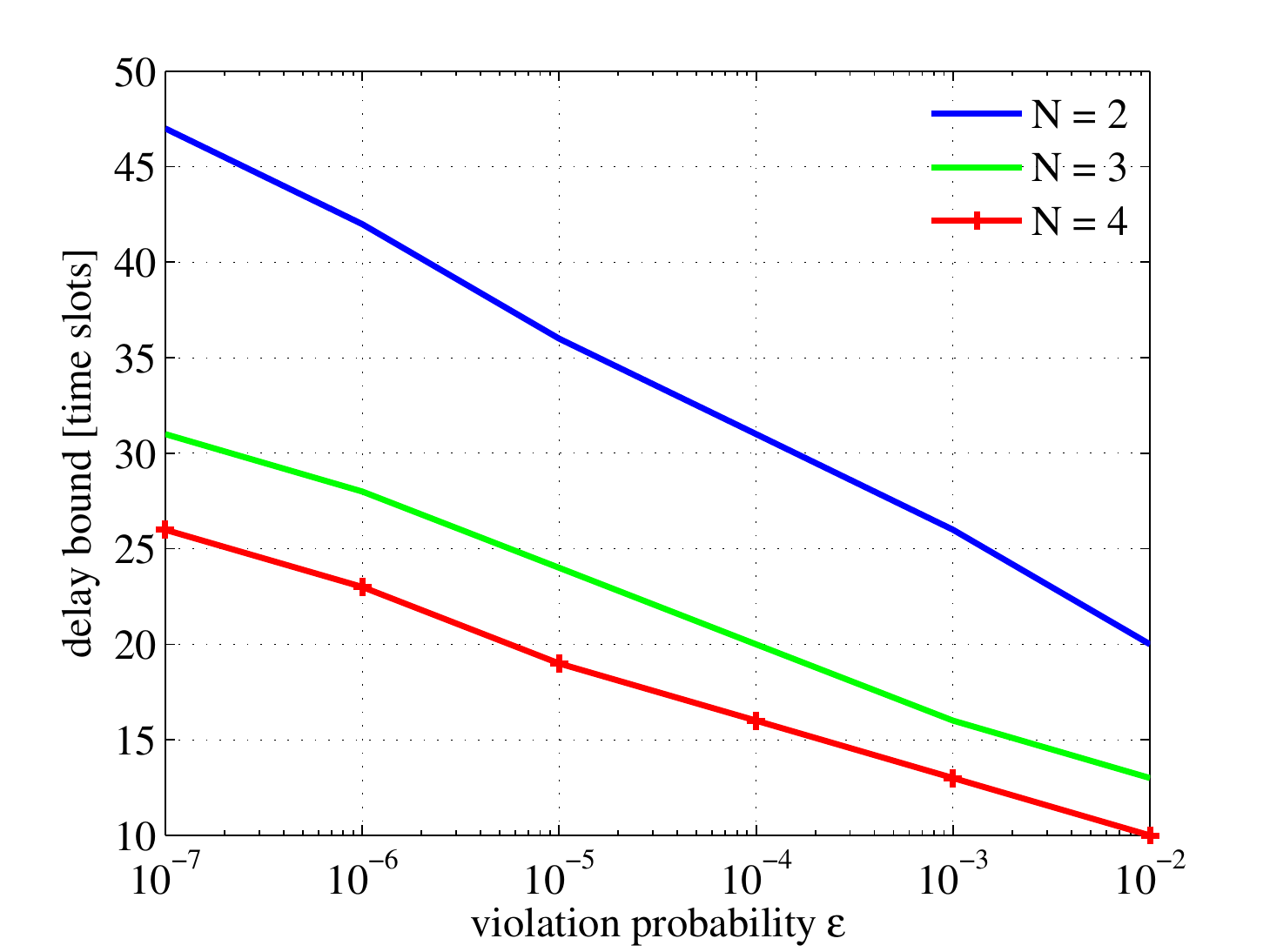}
	\caption{Stochastic delay bounds for different number of antennas N }
	\label{fig:dly_bnd_diff_nr_antennas}
\end{figure}
As discussed above an increase in the number of antennas leads in the best case to a linear increase in channel capacity. This increase capacity manifests itself in the decrease of the delay bound for fixed scenario conditions. This can be seen in Fig. \ref{fig:dly_bnd_diff_nr_antennas}. Further, observe the exponential tail decay of the delay bound. This is expected for the delay bound as it is derived using the MGFs and Chernoff's theorem. The figure shows additionally the change in the exponential decay factor, seen in the slope of the lines, which is introduced through the increase in the number of antennas N, respectively, capacity.

We next study the effect of fading speed on the delay bound.
We start from the configuration of transition probabilities $p_{bg}$ and $p_{gb}$ given before and fix the corresponding block error probability $\omega$ for each spatial path as given by \eqref{eq:block_error_prob}. We control the fading speed, under fixed $\omega$, by the mean time for the Markov chain of the path to change states twice.
To this end we variate $p_{bg}$ such that a small value corresponds to slow fading and vice versa.

Tab. \ref{tab:fading_meancap} verifies the known notion that higher antenna configuration provide a higher capacity.
Note that the ratio of the first order capacities of N = 3 to N = 2 stay unchanged over different fading speeds.
%\begin{figure}[t]
%	\centering
%		\includegraphics[width=0.90\columnwidth]{./fig/AvgCap_n2_n3_s4_p.pdf}
%	\caption{Impact of fading speed on the mean capacity}
%	\label{fig:fspeedAvgCap_n2_n3}
%\end{figure}
\begin{table}
\begin{center}
\caption{First order capacity for different \# of antennas}
\label{tab:fading_meancap}
\begin{tabular}{|c||c|} \hline
N & C [bps/Hz] \\ \hline
2 & 7.25 \\ \hline
3 & 10.5 \\ \hline
\end{tabular}
\end{center}
\end{table}

\begin{figure}[t]
	\centering
		\includegraphics[width=0.90\columnwidth]{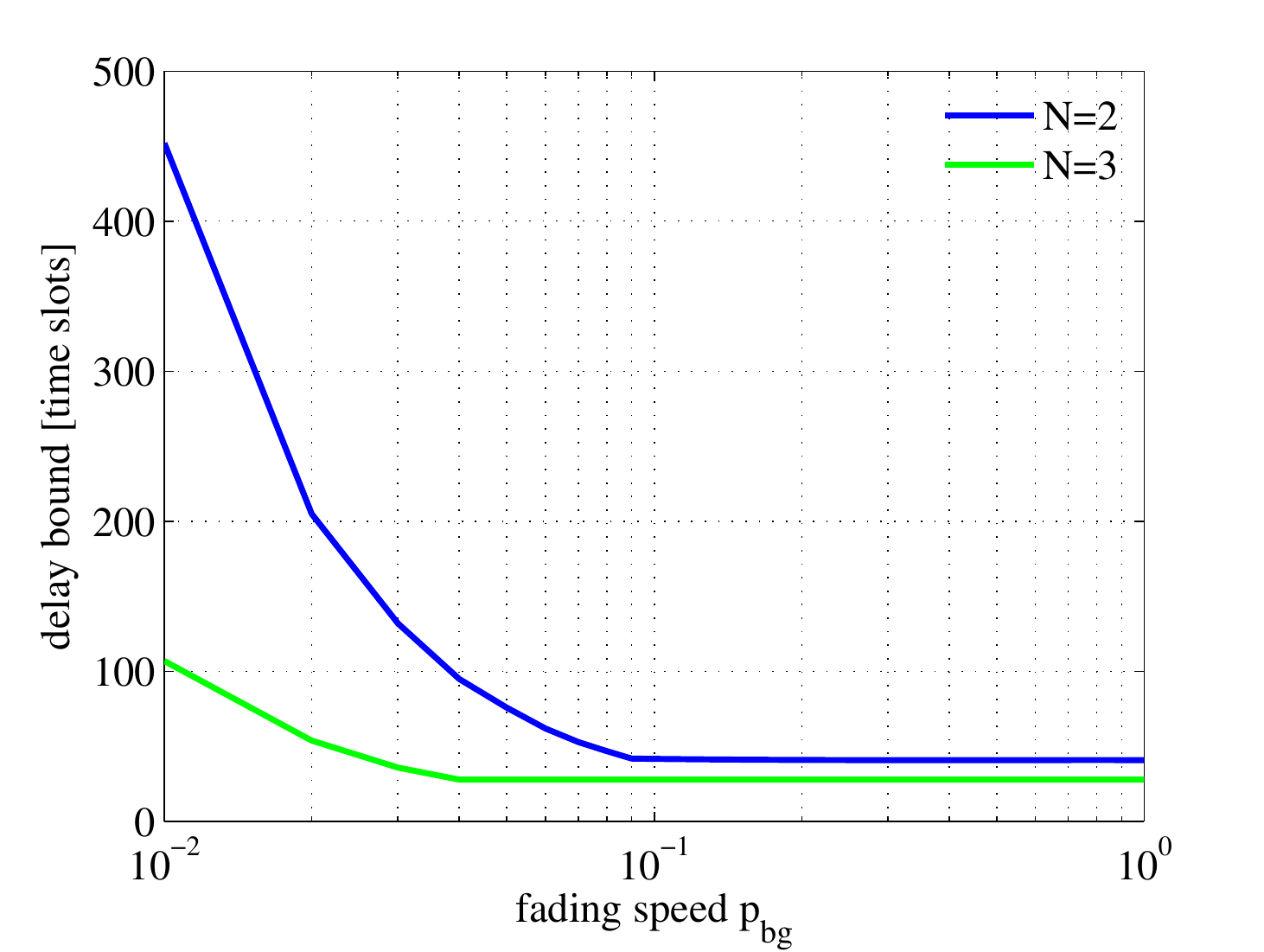}
	\caption{Impact of fading speed on the delay bound}
	\label{fig:fspeedDelay}
\end{figure}

The impact of the fading speed on the delay bound for different number of antennas (N = 2,3) is depicted in Fig.~\ref{fig:fspeedDelay}.
It shows that the delay bound deteriorates with slow fading significantly.
This is due to the fact that a slow fading channel, which corresponds to small values of $p_{bg}$, produces large burst errors due to its longer memory.
The notable thing is that for slow fading the difference in delay bounds between N = 2 and N = 3 is large and for higher fading speeds, which corresponds to a channel nearly without memory, the profit from extra spatial paths diminishes.
As the first order capacity given in Tab. \ref{tab:fading_meancap} stays constant over the range of fading speeds, which is implied by the fixed $\omega$, this difference in delay bounds might be due to the higher order statistics of capacity.
Furthermore, the impact of statistical multiplexing between the spatial paths, which is higher for $N = 3$, is apparent in the slow fading regime where the Markov chain changes states less frequently.

\begin{figure}[t]
	\centering
		\includegraphics[width=0.90\columnwidth]{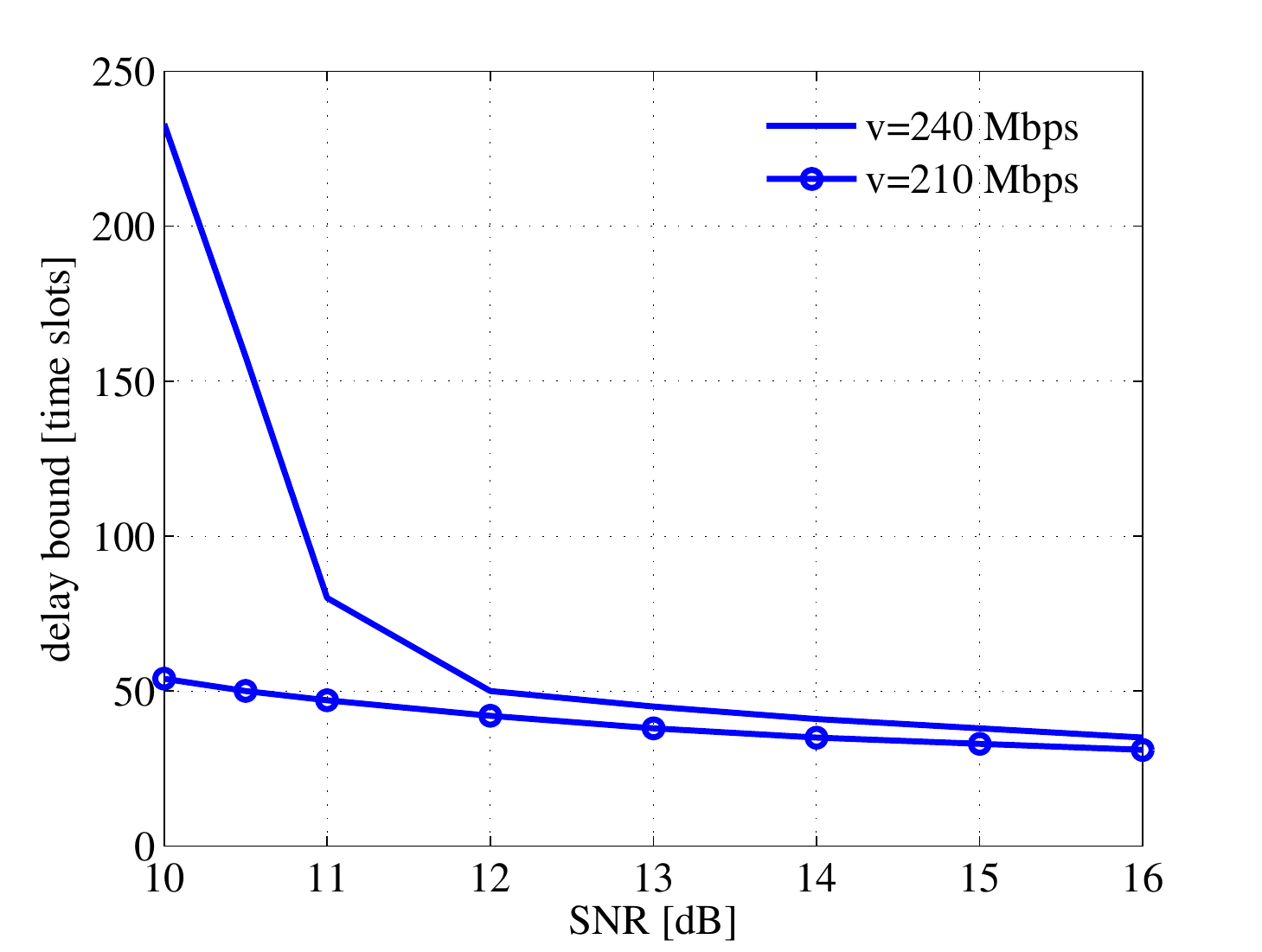}
	\caption{Impact of SNR on the delay bound for N = 2}
	\label{fig:dly_snr_N2}
\end{figure}

Fig.~\ref{fig:dly_snr_N2} shows results on the relation of the delay bound and SNR. One can observe a rapid growth of the delay bound occurring for small SNR values. This can be explained by the fact that the SNR governs the capacity expression (\ref{E:MIMOEPCapFormulalogDet}) such that for a constant arrival rate the decrease  in SNR corresponds to an increase in utilization, which, in turn for high values, causes large delays to occur more often and stochastic delay bounds to grow quickly.

\begin{figure}[t]
	\centering
		\includegraphics[width=0.90\columnwidth]{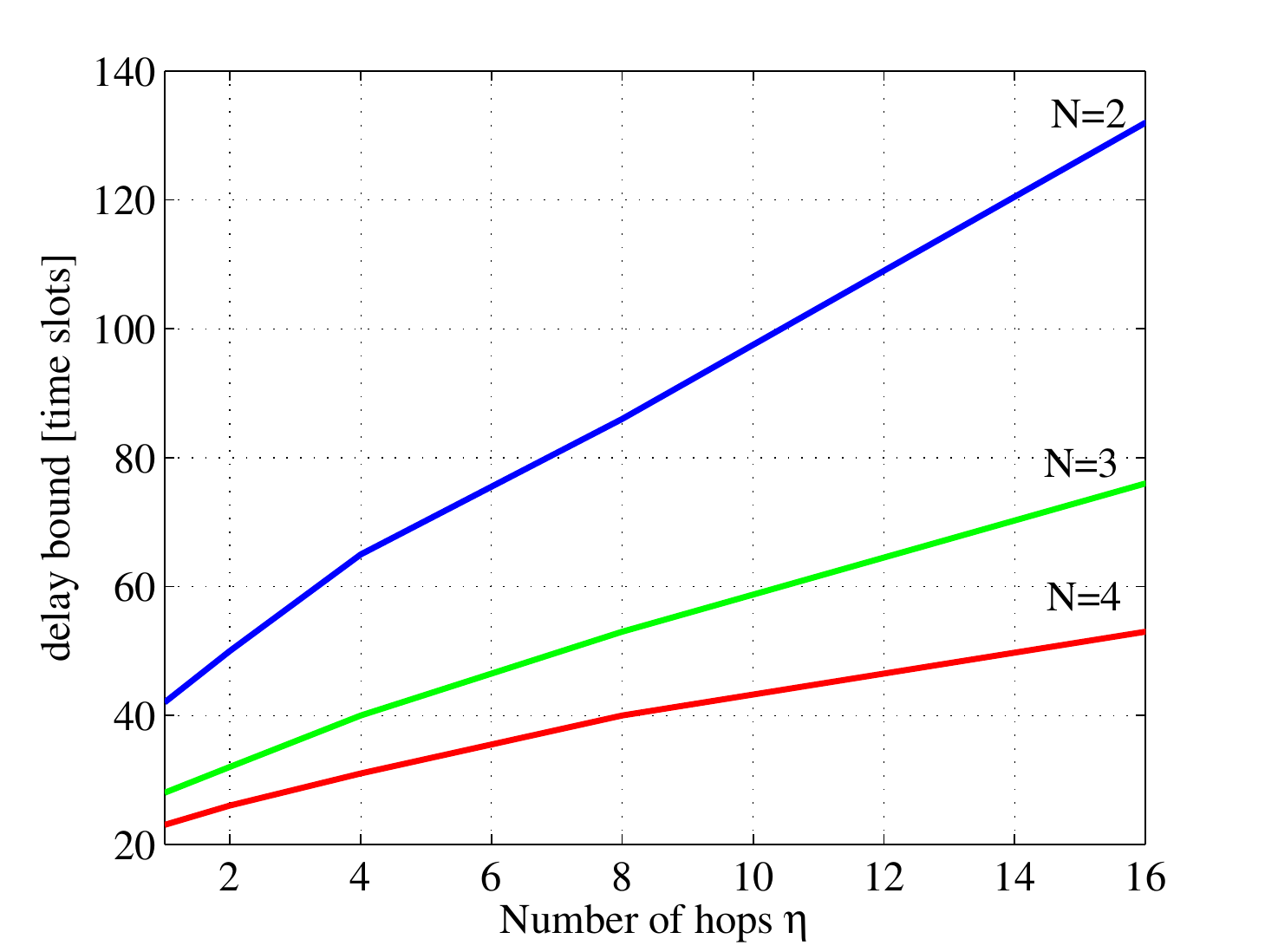}
	\caption{End-to-end delay bounds for different number of antennas on each hop.}
	\label{fig:dly_bnd_hops}
\end{figure}

%\begin{figure}
%\centering
%\subfigure[]{
%\includegraphics[width=0.90\columnwidth]{./fig/dly_n2_s45_p_eps6.pdf}
%\label{fig:n2_fspeed}
%}
%\subfigure[]{
%\includegraphics[width=0.90\columnwidth]{./fig/dly_n3_s45_p_eps6.pdf}
%\label{fig:n3_fspeed}
%}
%\caption{Impact of fading speed on the delay bound}
%\label{fig:fspeed}
%\end{figure}
%
%Figure \ref{fig:fspeed} shows the impact of the fading speed on the delay bound for different number of antennas (N=2,3). We set the block error probability for a sub-channel to be constant and control the fading speed by the mean time for the Markov chain of the sub-channel to change states twice. To this end we variate $p_{bg}$ such that a small value corresponds to slow fading and vice versa. Both subfigures \ref{fig:n2_fspeed}, \ref{fig:n3_fspeed} show that the delay bound deteriorates with slow fading significantly. This is due to the fact that a slow fading channel produces large burst errors due to its longer memory. In addition, the difference between both subfigures which can be at the scale of the ordinate is explained by the multiplexing gain of using multiple antennas.
%
%
For Fig.~\ref{fig:dly_bnd_hops}, we consider a simple multihop scenario, where we assume statistical independence of the service provided by the individual links. This scenario describes an ad-hoc wireless network, where we use homogeneous links for simplicity, i.e. all links are operated by $N$x$N$ MIMO in spatial multiplexing mode, here $N \in \{2,3,4\}$. Note that the framework described in Sect. \ref{sec:prelim} enables the derivation of delay bounds for a heterogeneous setting, by using the corresponding MGFs. Fig.~\ref{fig:dly_bnd_hops} depicts the growth of the probabilistic delay bound in the number of wireless hops.   For the MGF of the end-to-end service we use (\ref{eq:multihop}).  The first observation in Fig.~\ref{fig:dly_bnd_hops} is that the delay bounds scale at most linearly in the number of hops $\eta$. This observation is in accordance with the scaling of ${\mathcal O}\left(\eta\right)$ derived for MGF NetCal first in \cite{NetCal:Fidler2006:EoEProbabNetCalWithMGF}. The second observation is that the number of antennas $N$ changes the slope of the delay bound growth in the number of hops $\eta$. The additional capacity, which is provided by the allocation of additional antennas in the MIMO system, causes the change in the slope of the bound scaling.
\section{Conclusions and Future Work}
In this paper we conduct a flow-level performance analysis of a MIMO system with spatial multiplexing. The analytical framework we use is stochastic network calculus with moment generating functions. We model the block fading MIMO wireless channel as a Markov Chain and parameterize it to the application of IEEE 802.11n. The states of the Markov chain are decided on the degree of freedom, which results in a significant state space reduction from a direct way of defining the states. We show the improvement related to spatial multiplexing on the flow-level probabilistic delay bound. We also present numerical results of the delay bound and quantify the impact of increasing the number of antennas. Furthermore, we show results for end-to-end delay bounds in the independent multi-hop wireless scenario and the impact of increasing the number of antennas characterizing the MIMO links. While in this work periodic source has been used in illustrating the results, the same methodology can be applied to any traffic source with known MGF. For multi-user scenarios, it is left as a future work.

\end{document}